\documentclass[aps,prl,twocolumn,showpacs]{revtex4}
\usepackage[dvips]{graphicx}

\newcommand{\be}{\begin{equation}}
\newcommand{\ee}{\end{equation}}

\begin{document}

\title{History-Induced Critical Behavior in Disordered Systems}

\author{John H.\ Carpenter}
\author{Karin A.\ Dahmen}
\author{Andrea C.\ Mills}
\author{Michael B.\ Weissman}
\affiliation{
University of Illinois at Urbana-Champaign,
Department of Physics,
1110 West Green Street, 
Urbana, IL 61801.
}

\author{Andreas Berger}
\author{Olav Hellwig}
\affiliation{
Hitachi Global Storage Technologies,
San Jose Research Center,
650 Harry Road, E3
San Jose, CA 95120.
}

\date{\today}

\begin{abstract}
  
Barkhausen noise as found in magnets is studied both with and
without the presence of long-range (LR) demagnetizing fields using
the non-equilibrium, zero-temperature random-field Ising model. Two
distinct subloop behaviors arise and are shown to be in qualitative
agreement with experiments on thin film magnets and soft
ferromagnets.  With LR fields present subloops resemble a
self-organized critical system, while their absence results in
subloops that reflect the critical point seen in the saturation loop
as the system disorder is changed.  In the former case, power law
distributions of noise are found in subloops, while in the latter
case history-induced critical scaling is studied in avalanche size
distributions, spin-flip correlation functions, and finite-size
scaling of the second moments of the size distributions.  Results
are presented for simulations of over $10^8$ spins.

\end{abstract}


\maketitle



Hysteresis occurs when a system far from equilibrium is driven by an
external force. The system state then depends on the history of the
system.  In many systems such as ferromagnets \cite{Co91},
superconductors \cite{Fi95,Wu95}, and martensites \cite{ortin}, the
response to the driving force is not continuous but occurs in discrete
jumps which is often referred to as ``crackling noise'', or specifically
for magnets, Barkhausen noise.  In magnets, broad ranges of power-law
scaling of noise have been attributed to either a
disorder induced critical point, such as found in the non-equilibrium
zero temperature random-field Ising model (RFIM) \cite{hysterI}, or
to self-organized criticality (SOC) \cite{Zap98a}, as found in soft magnets.

The disorder induced critical point found in the saturation
magnetization curve, traced out as an external magnetic field drives
the system from one saturated state to the oppositely magnetized state
and back, has been studied in detail for the RFIM
\cite{Per95,Pe99,hysterI,thesis} and its existence has been
experimentally confirmed \cite{Ber00}.  However, taking magnets to
saturation often is impractical due to the large magnetic fields
required, so the behavior of subloops is of great interest to
experiments and applications alike.  The RFIM may be used to model
subloops, and magnetization curves have been computed exactly in one
dimension \cite{Sh00} and on a Bethe lattice \cite{Sh01} and have also
been collapsed near the demagnetized state using Rayleigh's Law
\cite{Zapperi1}.  Also, the idea of history acting as a source of
effective disorder was recently introduced \cite{Car01}. In this
letter we report how the presence or lack of long range (LR)
demagnetizing fields produces two distinct behaviors in subloops of
the RFIM. In particular, in the presence of sufficiently large LR
fields, subloops are found to resemble a SOC system. However, in the
absence of LR fields, we find a {\it critical subloop} inside the
saturation loop where the system history acts as a tuning parameter
instead of the system disorder.  Correspondingly, a modified scaling
picture for subloops, consistent with the saturation loop, is
introduced.  For both cases we report new experimental results for
subloops showing qualitative agreement with the two behaviors of the
RFIM.


In the non-equilibrium, zero-temperature RFIM spins $s_i = \pm 1$ are
placed on a hyper-cubic lattice.  At each site $i$ a quenched random local
field $h_i$ is chosen from a Gaussian distribution, $\rho(h_i) =
\frac{1}{\sqrt{2 \pi}R} exp(-\frac{h_i^2}{2 R^2})$.  The field $h_i$
acts as a source of disorder for the system, and the standard
deviation $R$ of the random distribution is termed the 'disorder'.
The energy of a system with N spins is given by \cite{Ku01}
\be
{\mathcal H} = -J\sum_{\langle ij \rangle}s_i s_j - \sum_i (h_i + H)
s_i + \frac{J_{LR}}{N} \sum_{ij} s_i s_j,
\label{Hamiltonian}
\ee
where $H$ is a tunable external magnetic field. The first term of
Eq.\ \ref{Hamiltonian} couples nearest-neighbor spins ferromagnetically
($\langle ij \rangle$ implies summing over nearest-neighbor pairs of
spins) while the last term provides an infinite range
anti-ferromagnetic (AF) coupling, where one sums over all pairs of
spins regardless of their relative distance. This AF coupling models
the dipolar interactions relevant in soft ferromagnets,
which Zapperi, et.~al.\ \cite{Zap98a,Zap98b} have shown in three dimensions to
have the same effect on long length scales as infinite range mean
field interactions.  Note that by choosing $J_{LR} = 0$ one recovers
the nearest-neighbor RFIM. In order to model hysteresis we study the
model at zero temperature, far from thermal equilibrium \cite{Pe99,hysterI}, and for convenience set $J=1$.

Simulations of the above model were run by starting with the external
field at $H = -\infty$ with all spins down, and then adiabatically
slowly moving the external field through a particular history. As the
field $H$ is changed, a given spin will flip (either upwards or
downwards) when its effective local field,
$h_i^{eff} = H + h_i + J\sum_{\langle ij \rangle} s_j -J_{LR} M$,
changes sign. Here $M = \frac{1}{N}\sum_i s_i$ is the magnetization of
a system with $N$ spins.  When a spin flips, it may induce its
neighboring (or for $J_{LR} \not = 0$ even distant) spins to flip,
creating an avalanche of flipping spins, which is the analog of a
Barkhausen pulse. The simulations are based on the code available on
the web \cite{web} which has been modified to allow for subloops in the
history. The code uses the sorted list algorithm, which stores the
random fields, and is described in depth in Ref.\ \cite{Ku99}.

In the absence of LR fields, as one tunes the system disorder ($R$),
the RFIM exhibits a non-equilibrium second order phase transition at a
critical disorder $R = R_c$ \cite{hysterI}. Below $R_c$ the coupling
of nearest-neighbors dominate the dynamics, and there is a finite jump
in magnetization of the saturation loop.  However, above $R_c$ the
random fields dominate, resulting in smooth hysteresis curves and
mostly small avalanches. Many quantities associated with this critical
point display scaling behavior for $R \rightarrow R_c$. A detailed
discussion is given in Ref.\ \cite{Pe99}.

\begin{figure}
\includegraphics[angle=0,width=8.2cm]{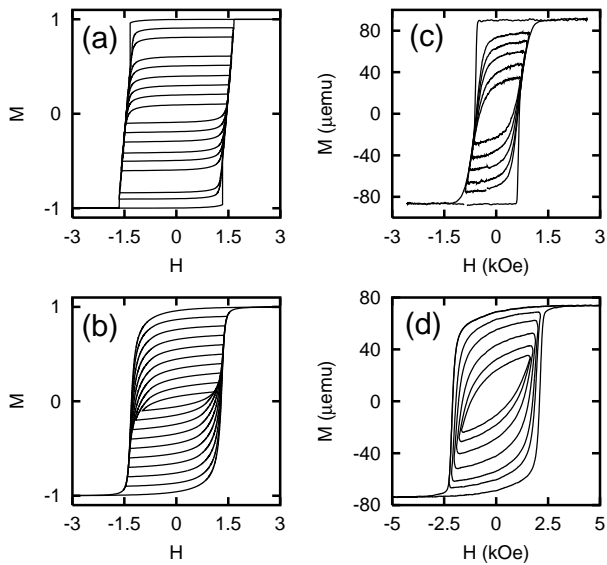}
\caption[Hysteresis loops] {
\label{mhfig}
Hysteresis loops with concentric inner subloops for (a) RFIM with
$240^3$ spins, $J_{LR}= 0.25$, and $R = 1.8$, (b) RFIM with $240^3$
spins, $J_{LR}= 0$, and $R=2.7$, (c) a Co/Pt multilayer thin
film, and (d) a CoPtCrB alloy thin film. 
}
\end{figure}

The two distinct behaviors of subloops in the RFIM can be seen in
Figure \ref{mhfig}.
In Fig.\ \ref{mhfig}(a) subloops spaced by $\Delta M = 0.1$ are shown
for a system with $J_{LR} = 0.25$ and a disorder of $R=1.8$, chosen
smaller than $R_c = 2.16$ as many experimental samples such as soft
ferromagnets are believed to be below the critical disorder. The
subloops show a linear $M(H)$ behavior as the field is increased, as
one would expect for a system undergoing domain wall propagation while
being exposed to a demagnetizing effect.  Only the first few subloops
show some effects of the initial condition (all spins down) before the
sweeping field has created a large enough domain to allow for single
domain wall propagation. On the other hand one finds a very different
picture in the absence of LR fields. In Fig.\ \ref{mhfig}(b), obtained
for a system with $J_{LR} = 0$, $R = 2.7$ and subloop spacing $\Delta
M = 0.1$, as one moves inwards, subloops begin to resemble the
saturation loop at an effectively higher disorder,
i.e. $R>2.7$. Indeed, pre-flipped spins not participating in a given
subloop may act as an added, possibly correlated, ``effective
disorder.'' Tuning the history in this way, we were unable to directly
observe a transition from loops with a jump in magnetization to smooth
inner subloops. This is due to the inability to break up the system
spanning avalanche present below the critical disorder even for the
largest simulated system sizes ($480^3$ spins)
\cite{Zapperi1,mckunpub,Car01}.

Experimental magnetization curves for thin films also exhibit these
two types of behavior depending on the presence of LR fields. Fig.\
\ref{mhfig}(c) displays magnetization curves for a Co/Pt-multilayer
with the field applied along the surface normal. Due to the strong
interface anisotropy of such multilayer samples, the easy axis of
magnetization is perpendicular to the film plane even though the LR
demagnetizing effect is largest in this direction.  Despite the fairly
rectangular major loop shape, indicating easy axis orientation, the
loop exhibits an extended linear segment on which all minor loops
merge due to the presence of LR dipolar effects. On the other hand,
films with in-plane magnetization behave quite differently, because LR
demagnetizing effects are extremely small in this geometry. Fig.\
\ref{mhfig}(d) shows magnetization curves for a CoPtCrB-alloy film
with the field applied in the film plane.  This film is
polycrystalline with grain sizes narrowly distributed around 10 nm
diameter and exhibits strong exchange coupling between grains. Due to
the lack of LR coupling, subsequent minor loops appear increasingly
sheared with decreasing coercive field, very similar to the curves
shown in Fig.\ \ref{mhfig}(b) for the $J_{LR} = 0$ RFIM. All
experimental data were measured at room temperature using an
Alternating Gradient Magnetometer.

\begin{figure}
\includegraphics[angle=0,width=8.3cm]{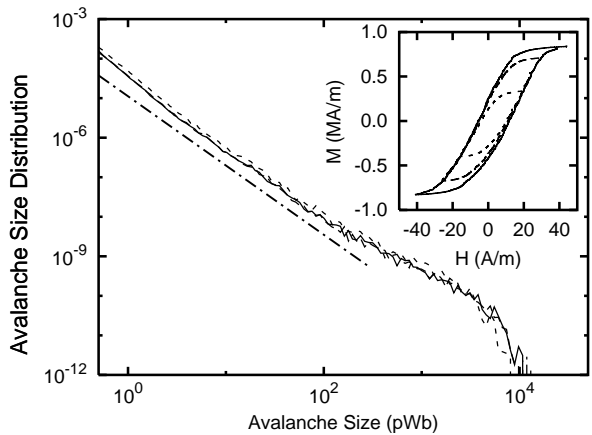}
\caption[Experimental avalanche size distribution] {
\label{exppsfig}
Experimental avalanche size distributions for subloops of a soft
ferromagnet. The three subloops analyzed are shown in the inset, with
the largest corresponding to the saturation loop. The
distributions were extracted from a window of width
$\Delta H \approx 20 A/m$ which started near $H = 0 A/m$. A power law
of $-1.75$ is show by the offset, dash-dot line.
}
\end{figure}

Much information on these two behaviors can be obtained by examining
the Barkhausen noise present in the system. With LR fields present,
the subloops display a power-law avalanche size
distribution, all with the same exponent and cutoff size, indicating that the system
is SOC \cite{Ku01,Zap98a,Zap98b}. Figure \ref{exppsfig}
displays the avalanche size distribution for subloops of a 21 cm x 1 cm x 30
$\mu$m ribbon of a Fe$_{21}$Co$_{64}$B$_{15}$ amorphous alloy. In the
experiment, a solenoid provides a triangular
driving field along the long axis of the sample at 0.03 Oe/s.
The first cycle drives the sample to saturation, while subsequent cycles
drive the field to successively smaller amplitudes (subloops).  The
driving rate
is slow enough to ensure that avalanches do not overlap.
The Barkhausen noise was measured with a pick-up coil of $150$ turns of
copper wire wound around the middle 0.3 cm of the sample.
By integrating the pick-up voltage one may obtain the magnetization. 
The resulting hysteresis loops are shown in
the inset of Fig.\ \ref{exppsfig}. The avalanche distributions for
the subloops are all identical, affirming the SOC behavior. The
measured power law exponent ($1.75$) is larger than has been predicted
and measured for the saturation loop ($1.25-1.5$) \cite{Ku01,Zap98b}.
This is due to
$M(H)$ not being strictly linear in the window analyzed. Choosing a
smaller window close to the linear regime results in a power law
exponent of $1.3$, consistent with previous measurements. The RFIM
with LR fields
presents an almost identical picture of subloops. For a $240^3$ system
with $R = 1.8$ and $J_{LR} = 0.25$, subloops displayed identical power laws
in their avalanche size distributions with a loop integrated exponent of $1.7 \pm 0.2$, in good
agreement with the experimental results.

In the absence of LR fields the case is much different; the system
history acts as a tuning parameter affecting a subloop dependent
cutoff. Although a transition like the one found in the saturation
loop cannot be directly observed, its existence may be ascertained by
examining the scaling behavior on one side of (``above'') the critical
point \cite{hysterI,Pe99}.  Thus we present a scaling collapse of loop
integrated spin-flip correlation functions and a finite-size scaling
collapse of the second moments of the integrated avalanche size
distribution for subloops with $J_{LR} = 0$.


For the spin-flip correlation function, in analogy with the
saturation loop \cite{Pe99} one may assume a scaling form
\be
\label{eq:genacfscale}
G^{int}(x,R,H_{max}) \sim x^{-(d+\beta_l/\nu_l)} {\mathcal
G}^{int}(x(R-R_c^d)^{\nu_l},xf^{\nu_f}), 
\ee
where $H_{max}$ is the maximum field along the subloop and
$f=H_{max_c}-H_{max}$ is the history-induced analog of the reduced
saturation loop disorder \footnote{Ref.\ \cite{Car01} used $\epsilon =
(M_{max_c}-M_{max})/M_{max}$ for the history 'disorder'.  However, $f$
provides a sounder scaling picture.}. Here $R_c^d$ is the critical
disorder for the demagnetization curve which is numerically within the
error bars of $R_c$. Similarly, $H_{max_c} = H_c^d$ is the critical
field associated with the demagnetization curve
\cite{Car03}. According to scaling theory one expects the exponents
$d+\beta_l/\nu_l$, $\nu_l$, and $\nu_f$ as well as the scaling
function ${\mathcal G}^{int}$ to be universal. Here a subscript $l$
denotes an exponent associated with the random disorder scaling of the
critical subloop and subscript $f$ one associated with the
history-induced ``disorder".  For the collapses, the system is run at
the effective critical disorder of the saturation loop, $R_c(L)$, for
the linear system size $L$ with $R_c(L) \rightarrow R_c$ as $L
\rightarrow \infty$. This system size dependent critical disorder
$R_c(L)$ is defined as the disorder at which the maximum number of
system spanning avalanches in the saturation loop is observed
\cite{Pe99}.

\begin{figure}
\includegraphics[angle=0,width=8.6cm]{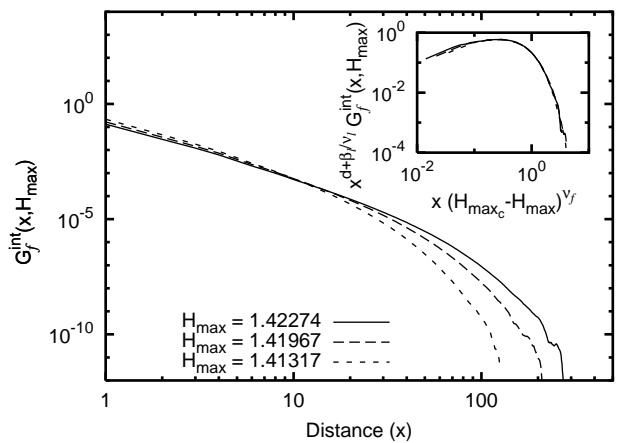}
\caption[Correlation function] { 
\label{acffig}
Integrated spin-flip correlation functions for $480^3$ systems at $R =
2.198$ and averaged over $20$ random seeds. Curves are given for
subloops starting at values of $H_{max} = 1.42274$, $1.41967$, and
$1.41317$. The inset contains a collapse of the three respective
distributions, yielding $d+\beta_l/\nu_l = 3.0 \pm 0.2$ and $1/\nu_f =
1.28 \pm 0.40$ with $H_{max_c} = 1.427$.
}
\end{figure}

The integrated spin-flip correlation function,
$G^{int}_f(x,H_{max})$, was measured for subloops spaced by $\Delta
M_{max} = 0.025$ in their maximum magnetizations (see Fig.~\ref{acffig}).
A similar behavior to the saturation loop is found. As the
history-induced disorder is increased, by moving to inner subloops,
the curves display a decreasing cutoff. As the system is started at
the effective critical disorder, Eq.~\ref{eq:genacfscale} reduces to 
$G^{int}_f(x,H_{max}) \sim x^{-(d+\beta_l/\nu_l)} {\mathcal
  G}^{int}_f(xf^{\nu_f})$.
This scaling ansatz results in the collapse shown in the inset of
Fig.~\ref{acffig}. Only subloops void of
spanning avalanches were used in the collapses to remove the effects
of the finite system size, equal to $480^3$ spins.

The integrated avalanche size distribution, with a similar scaling ansatz
$D^{int}_f(S,H_{max}) \sim S^{-(\tau + \sigma\beta\delta)_l} {\mathcal
  D}^{int}_f(S^{\sigma_f}f)$, was previously analyzed in
Ref.~\cite{Car01}. The conclusions therein remain valid, and the exponents
(see Table \ref{exptab}) have been updated to correspond with the use
of the scaling variable $f$.
\begin{table}
\caption[Critical Exponents] {
\label{exptab}
Universal critical exponents from scaling collapses in three
dimensions for the
history-induced
disorder present in subloops and the 
random disorder of the saturation loop.
}
\begin{ruledtabular}
\begin{tabular}{c c c}
Exponent & History-Induced & Saturation\footnote{References \cite{Pe99,Per95}} \\
\hline
$\tau+\sigma\beta\delta$ & $2.01 \pm 0.10$ & $2.03 \pm 0.03$ \\
$1/\sigma$               & $2.3 \pm 0.5$   & $4.2 \pm 0.3$ \\
$d + \beta/\nu$          & $3.0 \pm 0.2$   & $3.07 \pm 0.30$ \\
$1/\nu$                  & $1.28 \pm 0.40$ & $0.71 \pm 0.09$ \\
$\rho$                   & $2.6 \pm 0.40$  & $2.90 \pm 0.16$

\end{tabular}
\end{ruledtabular}
\end{table}

Table \ref{exptab} lists the exponents from the above two collapses
along with the saturation loop values.  While both power law
behaviors, given by $\tau+\sigma\beta\delta$ and $d+\beta/\nu$, are
identical within error, the exponents governing the cutoff behavior,
$\sigma_f$ and $\nu_f$, are found to differ from their saturation loop
counterparts with only a slight overlap in their error estimates. One
would expect identical values if the pre-flipped spins at the start of
a subloop were randomly distributed about the lattice, thus preserving
the uncorrelated, random nature of the system's disorder.  However,
this is not the case, as avalanches have left pockets of unflipped or
preflipped spins of all sizes up to the system's correlation length.
So the difference in exponents is not surprising as the
history-induced disorder acts more like a correlated disorder as
opposed to the uncorrelated, random system disorder $R$.

The second moments of the integrated avalanche size distribution also
behave similarly to the saturation loop and have a scaling form, in
analogy to the saturation loop and Eq.\ \ref{eq:genacfscale}, $\langle
S^2\rangle^{int}_f \sim L^{-\rho_l} {\mathcal S}^2_f(f L^{1/\nu_f})$
where $\rho_l = -(\tau_l+\sigma_l\beta_l\delta_l-3)/\sigma_l\nu_l$.
Table \ref{exptab} contains the exponents from a collapse for systems
with sizes from $L=80$ to $L=480$. As each system size was run at the
effective disorder, $R_c(L)$, a system size dependent $H_{max_c}(L)$
was required in the collapse.  The exponent $\nu_f$ was chosen to be
consistent across all collapses of the correlation function and second
moments. Its value differs from the value of $\nu$ for the saturation
loop, which reaffirms the differences between the history-induced and
random disorder.


Two new exponents, $\sigma_f$ and $\nu_f$, were introduced to describe
the history-induced scaling. However, only one is an independent
exponent as they obey the exponent relation $\sigma_f\nu_f=\sigma_l\nu_l$,
which may be derived from the relations in Ref.~\cite{Car03}.
Additionally, the critical subloop exponents may be close
to those of the saturation loop. Indeed, within error bars, numerical
results have found equal power-law exponents. However, currently no
exponent relations are known that require equality between the
saturation loop and subloop (subscript $l$) exponents.


The RFIM has been used to investigate subloops of the main saturation
hysteresis loop. Experimental measurement of subloops on thin film
magnets both with and without LR forces showed qualitative agreement
with the RFIM. In the presence of LR forces, subloops may be explained
by simple domain wall propagation, as confirmed by Barkhausen noise
measurements on soft magnets. However, in the absence of such forces
the disordered critical point was reflected in subloops. Scaling
collapses were performed for integrated avalanche size distributions,
integrated correlation functions, and the second moments of the
avalanche size distributions.  Finally, it was shown that only one of
the new critical exponents is independent of those found in the
critical subloop.

\begin{acknowledgments}
We thank Jim Sethna and Gary Friedman for very useful discussions
and Ferenc Pazmandi for suggesting to use $H_{max}$ as the
scaling variable instead of $M_{max}$. The work at UIUC was
supported by NSF grant DMRs 99-76550 (MCC), 00-72783, and 02-40644,
an A.\ P.\ Sloan fellowship (to K.\ D.), and an
equipment award from IBM.
\end{acknowledgments}

\end{document}